\documentclass[usenatbib,letters]{mnras}

\usepackage{newtxtext,newtxmath}
\usepackage[T1]{fontenc}

\usepackage{soul,xcolor}
\usepackage{cancel}
\DeclareMathAlphabet\mathbfcal{OMS}{cmsy}{b}{n}
\usepackage{bbold}
\usepackage{comment}
\usepackage[english]{babel}
\usepackage{amsmath}
\usepackage{lastpage}
\usepackage{graphicx}
\usepackage{xcolor}
\usepackage{comment}
\usepackage[version=4]{mhchem}
\usepackage{booktabs,siunitx}
\newcommand\mc[1]{\multicolumn{1}{c|}{#1}}
\usepackage{pifont}

\DeclareSIUnit\gauss{G}
\hypersetup{colorlinks=true, allcolors=blue}
\graphicspath{ {figures} }

\usepackage{ulem}

\newcommand{\Epar}{E_\parallel}
\newcommand{\rout}{r_{_{\rm out}}}
\newcommand{\rin}{r_{_{\rm in}}}
\newcommand{\bin}{\beta_{_{\rm in}}}
\newcommand{\bout}{\beta_{_{\rm out}}}
\newcommand{\me}{m_{\mathit{e}}}
\newcommand{\kb}{k_{\mathrm{B}}}
\newcommand{\tf}{t_{\mathit{f}}}
\newcommand{\tg}{t_{\mathit{g}}}
\newcommand{\rg}{r_{\mathit{g}}}
\newcommand{\Lbz}{L_{\rm BZ}}
\newcommand{\nel}{n_{\mathit{-}}}
\newcommand{\tmin}{\theta_{\rm{min}}}
\newcommand{\tmax}{\theta_{\rm{max}}}
\newcommand{\rh}{r_{_{\rm H}}}

\makeatletter
\newcommand{\fracto}[3]{%
  {\mathpalette\frac@to{{#1}{#2}{#3}}}%
}
\newcommand{\frac@to}[2]{\frac@@to#1#2}
\newcommand{\frac@@to}[4]{%
  \begingroup
  \sbox\z@{$\m@th#1\frac{#2}{#3}$}%
  \sbox\tw@{$\m@th#1\frac{#4}{#3}$}%
  \settowidth\dimen@{$\m@th\frac@to@demote#1#4$}%
  \frac{{}\makebox[\dimen@][l]{$\frac@to@demote#1#2$}}{#3}%
  \kern-\wd\tw@
  \kern\wd\z@
  \endgroup
}
\newcommand\frac@to@demote[1]{%
  \ifx#1\displaystyle\textstyle\else
  \ifx#1\textstyle\scriptstyle\else
  \scriptscriptstyle\fi\fi
}
\makeatother

\usepackage{mathtools}

\title[A kinetic study of black hole activation]{A kinetic study of black hole activation by local plasma injection into the inner magnetosphere}
\author[Niv et al.]{
Idan Niv,$^{1}$\thanks{E-mail: \href{mailto:idanniv@mail.tau.ac.il}{idanniv@mail.tau.ac.il}}
Omer Bromberg,$^{1}$
Amir Levinson,$^{1}$
Benoit Cerutti,$^{2}$
Benjamin Crinquand$^{3}$
\\
$^{1}$ The Raymond and Beverly Sackler, School of Physics and Astronomy, Tel Aviv University, Tel Aviv 69978, Israel\\
$^2$ Univ. Grenoble Alpes, CNRS, IPAG, 38000 Grenoble, France \\
$^3$  Department of Astrophysical Sciences, Peyton Hall, Princeton University, Princeton, NJ 08544, USA\\
}

\date{Accepted XXX. Received YYY; in original form ZZZ}

\pubyear{2023}

\begin{document}

\label{firstpage}
\pagerange{\pageref{firstpage}--\pageref{LastPage}}
\maketitle

\begin{abstract}
An issue of considerable interest in the theory of jet formation by the Blandford-Znajek mechanism, is how plasma is being continuously supplied 
to the magnetosphere to maintain it in a force-free state.  Injection of electron-positron pairs via annihilation of MeV photons, emitted from
a hot accretion flow, has been shown to be a viable possibility, but requires a high enough accretion rate.  At lower accretion rates, and in the 
absence of any other form of plasma supply, the magnetosphere becomes charge starved,
forming intermittent spark gaps that can induce intense pair cascades via interactions with soft  disk radiation, enabling outflow formation.
It is often speculated that enough plasma can penetrate the inner magnetosphere from the accretion flow through some rearrangement of magnetic
field lines (e.g., interchange instability).  However, the question arises whether such episodes of plasma intrusion 
can prevent the formation of spark gaps.  To address this question we conducted a suite of numerical experiments, by means of radiative, 2D axisymmetric general 
relativistic particle-in-cell simulations, in which plasma is injected into specified regions at a prescribed rate.  
We find that when pair production is switched off,  nearly complete screening is achieved when the plasma is injected within the outer light cylinder at a high enough rate.  
Injection beyond the outer light cylinder results in either, the formation of large vacuum gaps, or 
coherent, large-amplitude oscillations of the magnetosphere, depending on the injection rate.  Within the allowed dynamic range of our simulations, we
see no evidence for the system to approach a steady state as the injection rate is increased.  Switching on pair production results in 
nearly complete screening of the entire magnetosphere in all cases, with some fraction (a few percents) of the maximum Blandford-Znajek power emitted as TeV gamma-rays.
\end{abstract}


\section{Introduction}
A key issue in the theory of black hole (BH) outflows \citep{Blandford1977} is the nature of the plasma source in
the inner magnetosphere.   The activation of outflows by magnetic extraction requires continuous plasma production in the magnetospheric region enclosed between the inner and outer light surfaces, defined as the loci where the speed of an observer rotating with the magnetic flux tube equals the speed of light \citep{Blandford1977,Globus2013} \footnote{Formally these surfaces are the solutions to the equation $g_{\mu\nu}u^\mu u^\nu=0$, with $u^r=u^{\it\theta}=0$ and $u^{\it\phi} = \Omega\, u^{\it t}$ in Boyer-Lindquist coordinates, where $g_{\mu\nu}$ 
is the Kerr metric and $\Omega$ is  the angular velocity of magnetic field lines. It can be shown \citep{takahashi1990,Globus2013} that these are the surfaces on which the velocity of an ideal MHD flow equals the Alfv\'en velocity in the limit of zero inertia.}.   In order to establish a force-free jet, the plasma injection rate must be sufficiently high to maintain the density everywhere in the magnetosphere above a critical value, known as the Goldreich-Julian (GJ) density, \citep{Goldreich1969}.  If the plasma source cannot accommodate this requirement, charge starved regions (spark gaps) will
be created, potentially leading to self-sustained pair discharges.   In this scenario, charged leptons accelerated along magnetic field lines by the gap electric field scatter soft photons emitted by the surrounding matter to TeV energies. These gamma rays, in turn, interact with the soft photons to create more pairs, 
initiating pair cascades that tend to screen the gap, regulating the discharge process.   Analytic models \citep{Levinson2000,Neronov2007,Levinson2011,Hirotani2016} as well as general relativistic particle-in-cell (GRPIC) simulations \citep{Levinson2018,Chen2020,Crinquand2020,Crinquand2021,Kisaka2022} indicate that the energy dissipated in the 
gap is robustly emitted in the TeV band, and it has been 
speculated \citep{Levinson2000,Neronov2007,Levinson2011,Hirotani2016,Hirotani2016b,Levinson2018,KR2018,Chen2020,Kisaka2020,Kisaka2022} that 
this mechanism may explain the extreme TeV flares seen in M87 and, conceivably, other AGNs. 

A plausible plasma production mechanism that has been discussed extensively in the literature is annihilation of MeV photons emitted by the hot accretion flow (or a putative corona).
However, the pair injection rate predicted by this process is extremely sensitive to the rate at which plasma in the close vicinity of the BH is being accreted \citep{Levinson2011,moscibrodzka2011,Hirotani2016},
and a too low accretion rate is unable to produce enough plasma to continuously screen the magnetosphere everywhere. 
Whether this mechanism can provide complete screening of the BH magnetosphere in M87 is currently under debate \citep{Levinson2017}.  Here we
consider alternative injection processes that might operate in the absence of sufficient pair production opacity.
 %%f

One might speculate (as occasionally argued) that since the density of accreted plasma is much larger than the GJ density, 
screening of the magnetosphere by direct feeding of charges from the inner parts of the accretion flow might be viable.
Since the diffusion of charged particles across magnetic field lines is highly unlikely to supply sufficient plasma to the polar flow, given
that the cross-field diffusion time is vastly longer than the accretion time, one must resort to yet unspecified injection channel,
e.g., occasional rearrangement of magnetic surfaces at the jet boundary that might lead to sporadic loading of the inner magnetosphere.
To our knowledge, no such process has been identified in GRMHD simulations, however, one must keep in mind their limited resolution and dynamic range. 
But even if such episodic injections indeed occur in nature, it is
unlikely that plasma can be dumped continuously in the entire region encompassed 
between the inner and outer light surfaces.  The question is then how the magnetosphere of an active BH will respond to
injections in localized regions, for instance in the vicinity of the outer light surface.  
It could be that if the injected plasma is 
relativistically hot it quickly spreads over to cover the entire magnetosphere.  However, it is unclear whether the electric charge distribution 
imposed by the injection process will conspire to completely screen the magnetosphere. 
Alternatively, the inner magnetosphere will become highly intermittent in response to sporadic plasma injection. 
At any rate, if complete screening does not ensue, particles will 
be accelerated to high energies by the parallel electric fields generated in gaps ($\Epar= \mathbf{E}\cdot \mathbf{B}/B)$, 
producing pairs and high-energy radiation via interactions with soft photons emitted by the accretion flow, and via curvature radiation.

Motivated by the above consideration, we conducted a set of numerical experiments, by means of particle-in-cell (PIC) simulations, 
to explore how the magnetosphere responds to localized plasma injections.  Our experiments are restricted to steady injection 
in spherical shells (annuli in our 2D axisymmetric simulations).  We also conducted several experiments where injection is restricted to
a ring sector (in 2D) about the equatorial plane.  This configuration represents an accretion torus in more realistic situations.
%\benoit{Starting from here, this is a summary of this work findings and conclusion, not the right place for an introduction} 

Quite generally, we find that when plasma is injected in the entire causal region of the magnetosphere, 
complete screening ensues, even in the absence of external radiation, leading to 
the generation of a force-free outflow that appears to be in good agreement with the predictions of 
the Blandford-Znajek (BZ) mechanism.  However, in cases where the injection zone does 
not encompass the entire region between the inner and outer light surfaces and the interaction with disk radiation is switched off, 
a parallel electric field $\Epar$ is generated even when the injected plasma
is relativistically hot and the injection rate is relatively high (i.e., the mean pair density largely exceeds the GJ density in the injection zone).
The dynamics of the magnetosphere depends on the injection rate; when it is low enough (but still sufficiently high to maintain the
density in the injection zone well above the GJ density) a quasi steady state is established, whereby the amount of energy extracted from the black hole is small.
At higher injection rates the magnetosphere exhibits a cyclic dynamics, with (quasi) periodic modulations of the density and the parallel electric field
over a duration of tens $\tg$, resulting in the ejections of energy bursts with a maximum power 
that can reach $\sim 80$ percents of the optimal BZ power, $\Lbz$. 
When the interaction with disk radiation is switched on in these experiments, the system relaxes to a quasi steady force-free state,
with the extracted power reaching $\Lbz$, and the TeV luminosity of emitted radiation reaching a few percents $\Lbz$.

\section{Simulation setup}

We conducted 2D axisymmetric simulations with the PIC code Zeltron \citep{Cerutti2013}, modified to include GR effects \citep{Parfrey2019,Crinquand2020}. The system consists of a Kerr BH with a Kerr parameter $a=0.99$ threaded initially by a monopole magnetic field.  The choice of a monopole field was made to avoid the formation of current sheets at the equatorial plane, which complicate the analysis and the interpretation of the results.
We use geometrized units, where length scales and time are normalized by the BH gravitational radius, $\rg$ and $\tg=\rg/c$, respectively. 
Henceforth, densities are measured in units of a fiducial density, $n_0 = \Omega B_{\rm{H}}/2\pi e c$, where $B_{\rm{H}}$ is the magnetic field strength on the horizon,
$\Omega= \Omega_{\rm{H}}/2 $ is the angular velocity of the  monopole field, and $\Omega_{\rm{B}} = ac/2\rh\approx 1/2\tg$ is the BH angular velocity.
For this choice, the  associated plasma frequency is $\omega_{\mathit{p}} =\sqrt{\Omega_{\rm{H}} \omega_{\rm{H}}}$, with $\omega_{\rm{B}}= eB_{\rm{H}}/\me c$, the fiducial magnetization
is $\sigma_0 = \omega_B/ \Omega_{\rm{H}} \approx 2eB_{\rm{H}} \rg/\me c^2$, and the ratio of gravitational radius to skin depth is $\rg \omega_{\mathit{p}}/c = \sqrt{\sigma_0}/2$.
In M87 we typically have $\sigma_0\sim 10^{13}$. Such a value is unrealistic for GRPIC simulations that attempt to resolve the skin depth. In the simulations presented 
below we choose a rescaled value of $\sigma_0 = 5\times10^5$, which allows skin depth resolution in all cases studied (see \citealt{Crinquand2020} for further details).
It is worth noting that for the monopole field adopted here the magnetization at radius $r$ scales as $\sigma(r) \propto \kappa(r)^{-1}r^{-2} $, where 
$\kappa(r) = n(r)/n_0$ is the dimensionless pair density at radius $r$.

\begin{figure}
\includegraphics[width=\linewidth]{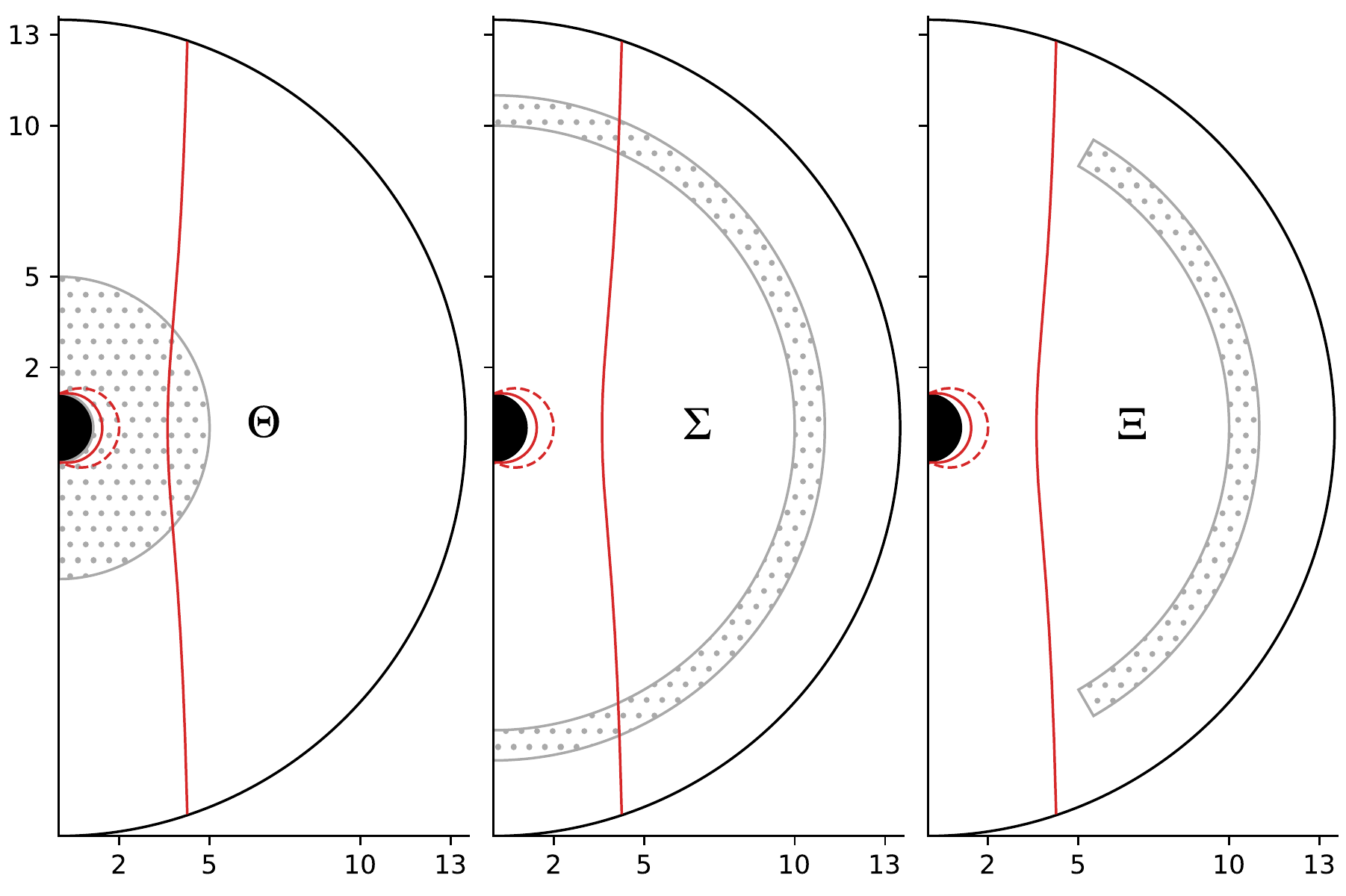}
\caption{The different types of models used in this work. The injection zones are marked with gray dots. The red solid lines mark the inner and outer light surfaces and the dashed line marks the outer surface of the ergosphere. The light surfaces are evaluated for a case of  $\Omega=\Omega_{\rm{H}}/2$.}
\label{fig:models}
\end{figure}

We used a grid of spherical Kerr-Schild coordinates that extends from $0.9 \rh$ to a radius of $15 \rg$, where we set an absorbing layer between $13.5-15~\rg$. Once the simulation starts we impose a steady injection of electron-positron pairs in a spherical shell between radii $\rin$ and $\rout$, where the pairs are distributed randomly inside the shell 
and have a thermal velocity distribution with a temperature $T$.
Table \ref{tab:models} shows the 3 types of models used in this work. Each simulation was run until it reached a steady state, or in cases where 
the system exhibited  cyclic dynamics (as in the models with high injection rate discussed below),
until it completed several cycles. The simulations were conducted in two limits. In the first we turned off 
Compton scattering (CS) decoupling the particles from the background radiation field. In this case particle flux is conserved outside the injection zone, while particles can exchange energy with the EM field and emit curvature radiation. In the second limit we turn on CS allowing for pair creation to take place in the box, which in turn allows for a more efficient screening of the parallel electric field reducing the energy gain from the EM field. 
We measured the Poynting flow and the energization of particles in the magnetosphere in each model and compared them to estimate its efficiency in activating the BH. 

\subsection{Electromagnetic fields}
In the $3+1$ formalism of \cite{Komissarov2004}, the electromagnetic tensor, $\mathit{F^{\mu\nu}}$, is decomposed into
electric field $\mathbf{D}$ and magnetic field $\mathbf{B}$, defined (in components) by
\begin{equation}
\mathrm{D}^\mathit{i}=\frac{1}{2}\varepsilon^{\mathit{ijk}}{{}^\ast\! F}_{\mathit{jk}},%\frac{1}{2\sqrt{\gamma}}
\end{equation}
and
\begin{equation}
\mathrm{B}^\mathit{i}=\frac{1}{2}\varepsilon^{\mathit{ijk}}F_{\mathit{jk}},%\frac{1}{2\sqrt{\gamma}}
\end{equation}
where $\mathit{{{}^\ast\! F}^{\mathit{\mu\nu}}}$ is the dual electromagnetic tensor,
$\gamma$ is the determinant of the three-dimensional metric tensor $\it \gamma_{ij}$ describing the space-like hypersurfaces in
the $3+1$ foliation, and $\varepsilon$ is its corresponding Levi-Civita tensor. 
The two general relativistic invariants can be expressed in terms of these fields as $\mathit{{{}^\ast\! F}_{\mu\nu}F^{\mu\nu}} = 4 \mathbf{D}\cdot\mathbf{B} $
and $\mathit{F_{\mu\nu}F^{\mu\nu}} = 2 ( \mathbf{B}^2 - \mathbf{D}^2 )$. In ideal MHD (or FFE) these invariants satisfy $\mathbf{D}\cdot\mathbf{B}=0$
and $\mathbf{B}^2 - \mathbf{D}^2 >0$.  In starved magnetospheric regions $\mathbf{D}\cdot\mathbf{B}\ne 0$.
Therefore, the quantity $\mathbf{D}\cdot\mathbf{B}/B^2$, which measures the strength of the electric field 
along magnetic field lines relative to the local magnetic field can be used to identify unscreened regions.

\subsection{Plasma injection scheme}
As explained above, in each numerical experiment pairs are injected in a spherical shell of inner radius $\rin$ and outer radius $\rout$.
The rate at which pairs are injected inside the shell is taken to be 
\begin{equation}
    \dot{n}_{\rm{inj}}=\dot{n}_0\frac{\rg^2}{r^2} = \chi n_0 c \frac{\rg }{r^2},
    \label{eq:injection_chi}
\end{equation}
where we adopt the normalization $n_0/\tg$, viz., $\dot{n}_0 =\chi n_0/\tg$ and $\chi$ is a dimensionless  factor.   
For the models listed in table \ref{tab:models}, the temperature of the injected plasma is mildly relativistic, $\kb T=\me c^2$, except for 
models $\tt \Sigma 5$, $\tt \Xi N$ and $\tt \Xi W$ for which it is ten times larger.  At such temperatures, the injected pairs should be able to propagate from 
the injection zone to other regions of the magnetosphere at nearly the speed of light. 

A rough estimate of the mean density 
in a shell far enough from the BH (where the metric is nearly flat) can be obtained upon assuming that the system is in a steady state and the density
inside the shell is uniform.  Equating the total rate of injection, 
$\int_{\rin}^{\rout} \dot{n}_{\rm{inj}}d^3r = 4\pi \chi n_0 c \rg(\rout- \rin)$,
with the rate at which plasma is lost from the shell boundaries, $4\pi n c (\rout^2\bout - \rin^2\bin)$, where $\bout>0 \, (\bin<0)$  is 
the radial bulk 3-velocity of the plasma escaping from the outer (inner) boundary,  one obtains:
\begin{equation}
n = \frac{\chi \rg(\rout-\rin)}{(\rout^2\bout - \rin^2\bin)} n_0.
\label{eq:injection}
\end{equation}
For the $\tt \Sigma$ models in table \ref{tab:models}, $\rout-\rin = \rg$, yielding 
$n \rout^2/n_0\rg^2 \approx \chi/(\bout - \bin)$.   
From the simulation we find $\bout - \bin \approx 0.3$, from which we obtain $n \rout^2/n_0\rg^2 \approx 3\chi$,
which is smaller by about a factor of 2 than the value measured in the simulation. 
For extended injection, with  $\rin \ll r$ and $\bout\approx 1$, we estimate the local density 
to be $n(r) \approx \chi n_0 (\rg/r) $ by setting $\rin=0$ and $\rout =r$ in Eq. (\ref{eq:injection}), 
or $n r^2/n_0 \rg^2 \approx \chi (r/\rg)$.  Thus, we generally anticipate the ratio between the density and the local GJ density to be 
of the order of a few times $\chi$, consistent with the results of the simulations.

\subsubsection{Photon generation and pair production}
In addition to the prescribed injection scheme described above, we also included in some of the runs photon generation
by inverse Compton scattering of disk radiation, and pair creation via  interactions of the IC gamma rays 
thereby produced with the same soft photons.  
Following  \citealt{Crinquand2020} 
we assume that the radiation field is time independent,
uniform, isotropic, and monoenergetic, with energy $\epsilon_0$ and
density $n_\text{soft}$. We do not include any feedback of the simulation
on this radiation field. The upscattered photons and created
leptons are assumed to propagate along the same direction as
their high-energy parents, reflecting strong relativistic beaming. 
The intensity of the background radiation field is quantified in table \ref{tab:models} by the fiducial optical depth 

\begin{equation}    \tau_0= \sigma_{\mathrm{T}} \rg n_\text{soft} \label{eq:tau},
\end{equation}
where $\sigma_\mathrm{T}$ is the Thomson cross section.  To guarantee optimum scale separation we adopt $\epsilon_0 = 5\times10^{-3}$ (see \citealt{Crinquand2020} 
for further details).

\begin{table*}
    \centering
    \begin{tabular}{c|r|r@{}l|c|r|r|c|r|c}
    \toprule
       Model  & $\tau_0$ & $\rin-$& $\rout$ & $\Delta\theta$ &\mc{$\chi$} & \mc{$\kb T / \me c^2$} & $\sigma_0$ & $\tf / \tg$ & Screening \\ 
        \midrule
\href{https://www.youtube.com/watch?v=4b-OPT23kJ0&list=PLsP9qR29TtwnAsBPx5xTPEUy6FZ_b-0PX&index=1}{$\tt \Theta S1$} & 0 & $\rh-$&$ 2$ & $\pi$ & 1 & 1 &  $5\times10^5$ & 99 & no \\[0.1 cm]
  %    \hline
\href{https://www.youtube.com/watch?v=gO3LTNjhpck&list=PLsP9qR29TtwnAsBPx5xTPEUy6FZ_b-0PX&index=2}{$\tt \Theta S2$} & 0 & $\rh-$&$ 2$ & $\pi$ & 5 & 1 & $5\times10^5$ & 140 & no \\[0.1 cm]
  %    \hline
\href{https://www.youtube.com/watch?v=m4IFhHi9I70&list=PLsP9qR29TtwnAsBPx5xTPEUy6FZ_b-0PX&index=3}{$\tt \Theta S3$} & 0 & $\rh-$&$ 2$ & $\pi$ & 10 & 1 & $5\times10^5$ & 99 & no \\[0.1 cm]
  %    \hline
\href{https://www.youtube.com/watch?v=5DY3XxRtCoA&list=PLsP9qR29TtwnAsBPx5xTPEUy6FZ_b-0PX&index=4}{$\tt \Theta I$} &  0 & $\rh-$&$5$ & $\pi$ & 1 & 1 & $5\times10^5$ & 99 & yes \\[0.1 cm]
%\hline
\href{https://www.youtube.com/watch?v=Q-Mbkoq1jqc&list=PLsP9qR29TtwnAsBPx5xTPEUy6FZ_b-0PX&index=5}{$\tt \Theta L$} & 0 & $\rh-$&$13.5$ & $\pi$ & 1 & 1 & $5\times10^5$ & 99 & yes \\[0.1 cm]
%\hline
\href{https://www.youtube.com/watch?v=i8Nt_DhlYog&list=PLsP9qR29TtwnAsBPx5xTPEUy6FZ_b-0PX&index=6}{$\tt \Sigma1$} &  0 & $10-$&$11$ & $\pi$ &  1 & 1 & $5\times10^5$ & 99 & no \\[0.1 cm]
%\hline
\href{https://www.youtube.com/watch?v=36wsDA1g90M&list=PLsP9qR29TtwnAsBPx5xTPEUy6FZ_b-0PX&index=7}{$\tt \Sigma2$} & 5 & $10-$&$11$ & $\pi$ & 1 & 1 & $5\times10^5$ & 99 & no 
 \\[0.1 cm]
\href{https://www.youtube.com/watch?v=LEVFKT4xrnY&list=PLsP9qR29TtwnAsBPx5xTPEUy6FZ_b-0PX&index=8}{$\tt \Sigma3$} & 10 & $10-$&$11$ & $\pi$ & 1 & 1 & $5\times10^5$ & 99 & partial \\[0.1 cm]
\href{https://www.youtube.com/watch?v=mVpgLdkj7Rs&list=PLsP9qR29TtwnAsBPx5xTPEUy6FZ_b-0PX&index=9}{$\tt \Sigma4$} & 20 & $10-$&$11$ & $\pi$ & 1 & 1 & $5\times10^5$ & 99 & yes \\[0.1 cm]
%$\mathcal{M}_{0,2}^\star$ & 0.0 & $r_h-$&$2$ & 5.47 & 1.00 & 0.350 & 99.22 \\
\href{https://www.youtube.com/watch?v=m-P7DIaDL1w&list=PLsP9qR29TtwnAsBPx5xTPEUy6FZ_b-0PX&index=10}{$\tt \Sigma5$} & 0 & $10-$&$11$ & $\pi$ & 10 & 10 & $5\times10^5$ & 163 & damped oscillations \\[0.1 cm]
\href{https://www.youtube.com/watch?v=nCQv_rRwaF0&list=PLsP9qR29TtwnAsBPx5xTPEUy6FZ_b-0PX&index=11}{$\tt \Sigma6$} & 0 & $10-$&$11$ & $\pi$ & 30 & 10 & $5\times10^5$ & 123 & periodic \\[0.1 cm]
\href{https://www.youtube.com/watch?v=_4GeHcvL-vQ&list=PLsP9qR29TtwnAsBPx5xTPEUy6FZ_b-0PX&index=12}{$\tt \Xi N1$} & 0 & $10-$&$11$ & $\pi/3$ & 30 & 10 & $5\times10^5$ & 99 & no \\[0.1 cm]
\href{https://www.youtube.com/watch?v=VaTN9rXJXsM&list=PLsP9qR29TtwnAsBPx5xTPEUy6FZ_b-0PX&index=13}{$\tt \Xi N2$} & 0 & $10-$&$11$ & $\pi/3$ & 300 & 10 & $5\times10^3$ & 37 & yes \\[0.1 cm]
\href{https://www.youtube.com/watch?v=xiMUd2gR91w&list=PLsP9qR29TtwnAsBPx5xTPEUy6FZ_b-0PX&index=14}{$\tt \Xi N3$} & 20 & $10-$&$11$ & $\pi/3$ & 30 & 10 & $5\times10^5$ & 99 & yes \\[0.1 cm]
\href{https://www.youtube.com/watch?v=bBPDAaQdki0&list=PLsP9qR29TtwnAsBPx5xTPEUy6FZ_b-0PX&index=15}{$\tt \Xi W1$} & 0 & $10-$&$11$ & $2\pi/3$ & 15 & 10 & $5\times10^5$ & 99 & no \\[0.1 cm]
\href{https://www.youtube.com/watch?v=DTPQhdFk0Vs&list=PLsP9qR29TtwnAsBPx5xTPEUy6FZ_b-0PX&index=16}{$\tt \Xi W2$} & 0 & $10-$&$11$ & $2\pi/3$ & 150 & 10 & $5\times10^3$ & 40 & yes \\[0.1 cm]
\href{https://www.youtube.com/watch?v=YheJuy4QhSE&list=PLsP9qR29TtwnAsBPx5xTPEUy6FZ_b-0PX&index=17}{$\tt \Xi W3$} & 20 & $10-$&$11$ & $2\pi/3$ & 15 & 10 & $5\times10^5$ & 34 & yes \\[0.1 cm]
        \bottomrule
    \end{tabular}
   \caption{A list of the models discussed in the text.  The corresponding configurations of the injection zone are presented in Fig. \ref{fig:models}.
  The models differ by their opacity for pair creation $\tau_0$ (Eq. \ref{eq:tau}), injection zone geometry, injection rate $\chi$ (Eq. \ref{eq:injection_chi}), 
  fiducial magnetization $\sigma_0$ and temperature of injected plasma.  In models $\tt \Xi N$ and $\tt \Xi W$ the injection zone is a ring sector of angular 
  width $\Delta\theta = \pi/3$ and $2\pi/3$, respectively (see Sec. \ref{sec:torus} for further details). Each model is linked to a movie that shows the time evolution of the electron number density ($\nel r^2/n_0 \rg^2$), parallel electric field ($\mathbf{D}\cdot\mathbf{B}/B^2$) and power from the BH $\left(\int \it{T^r_{\ t}} dA/\Lbz\right)$. The movies are accessible in the on-line version by pressing on the model name (in blue text).}
    \label{tab:models}
\end{table*}

\section{Results}
\begin{figure}
\includegraphics[width=\linewidth]{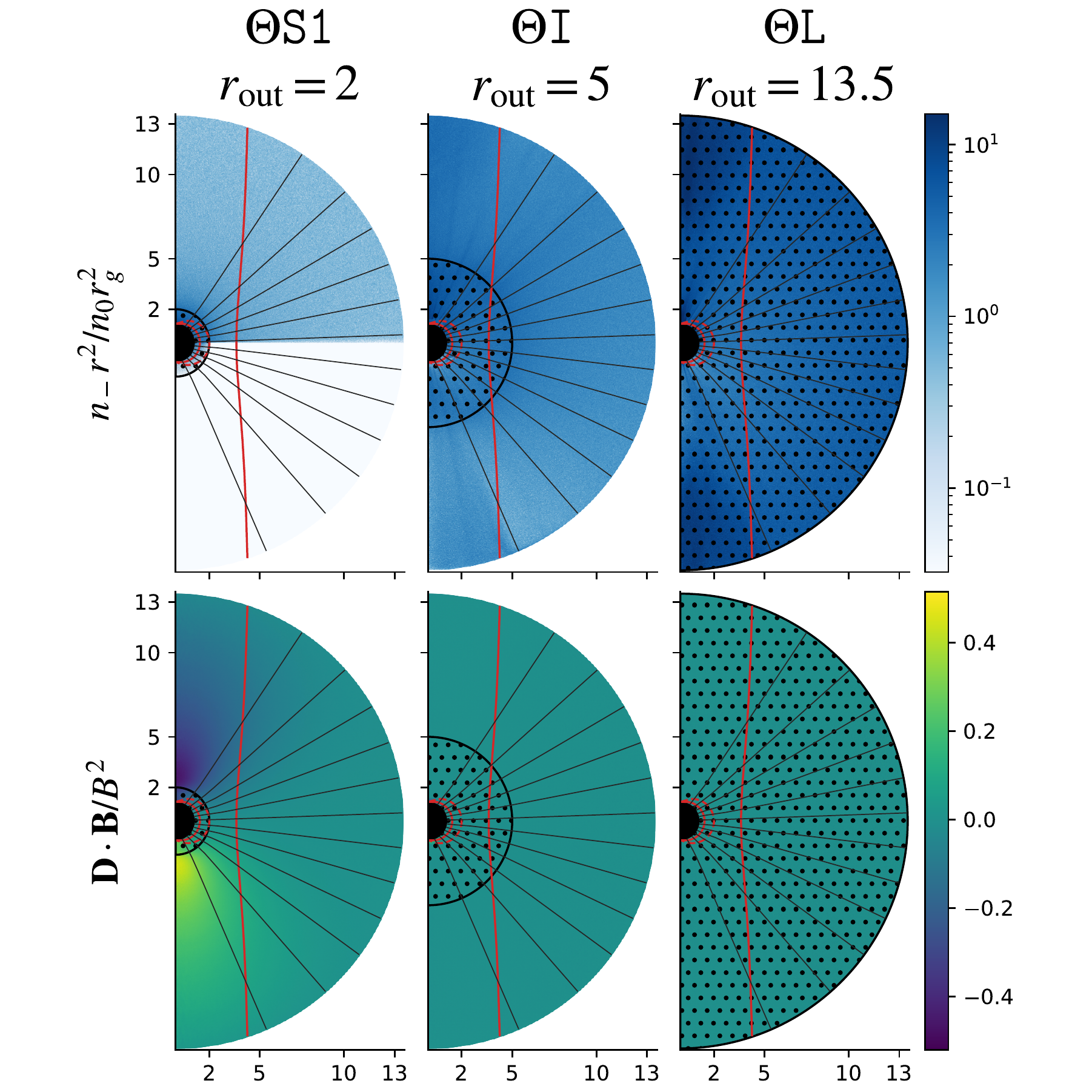}
\caption{Electrons number density (top) and normalized parallel electric field,
$\mathbf{D}\cdot \mathbf{B} /B^2$ (bottom), for cases of plasma injection between $\rin=\rh$ and (left to right) $\rout=2,5,13.5~\rg$. In all cases shown $\chi=1$. 
The injection zones are marked with black dots, magnetic field lines with gray solid lines and the inner and outer light surfaces with solid red lines. A nearly complete screening is obtained in the two right cases where the injection zone extends beyond the outer light cylinder. }
\label{fig:density_Epar_inner}
\end{figure}

In order to examine the effect of external plasma injection on the dynamics of the magnetosphere,
we run a series of models where we varied the size and location of the injection zone, the injection rate and the optical depth for photon-photon pair creation.  The different models are listed in table 1. 
%Each model is designated  by an upper index that indicates the value of the fiducial optical depth $\tau_0$ (Eq. \ref{eq:tau}), and a lower index marking the inner and outer radii of the injection ring in units of $\rg$. 
In what follows, cases in which the plasma injection zone encompasses the region below the outer light surface (left
panel in Fig. \ref{fig:models}) are termed "internal injection", otherwise they are termed  "external injection".

\subsection{Internal injection}
In the first suite of experiments we fixed $\rin$ at the BH horizon and varied $\rout$ (models $\tt \Theta S1$, $\tt \Theta I$, $\tt \Theta L$).  The interaction with the external radiation was switched off by setting $\tau_0=0$.
Each model was run for a long enough time to allow the system to reach a quasi steady-state (typically after about $30 \rg$).
Figure \ref{fig:density_Epar_inner} shows a comparison of the three models well after the system in each case has reached the quasi steady-state phase.
The top panels show the number density of electrons, $\nel$, in units of $n_0(\rg/r)^2$.
The distribution of positrons is a mirror image with respect to the $x$ axis and is not presented.
The bottom panels show the quantity $\mathbf{D}\cdot \mathbf{B}/B^2$, which indicates the level of charge starvation in magnetospheric zones.
As seen, effective screening of the entire magnetosphere is established in models $\tt \Theta I$ and $\tt \Theta L$, in which the injection zone extends beyond the outer light cylinder (marked with a solid red vertical line). In model $\tt \Theta S1$, wherein the plasma is injected within a radius of $r\leq2\rg$, a strong parallel electric field is generated in a large portion of the magnetosphere above and below the equatorial plane. 

\begin{figure}
\includegraphics[width=\linewidth]{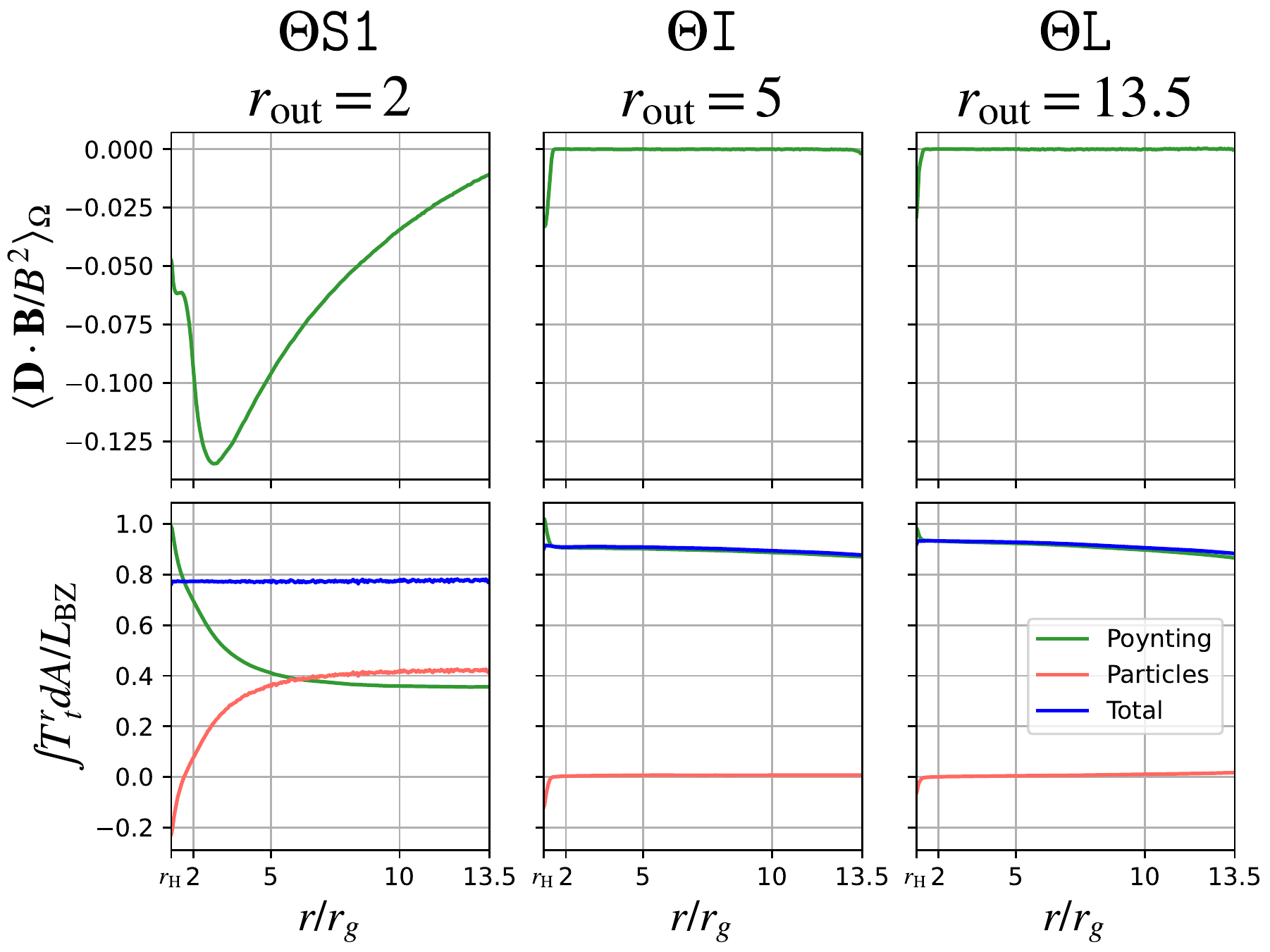}
\caption{The radial distribution of the solid angle-averaged northern hemisphere parallel electric field, $\langle\mathbf{D}\cdot\mathbf{B}/B^2\rangle_\Omega$ (top), and  normalized power $\int \it{T^r_{\ t}}dA/\Lbz$ (bottom), in cases with $\tau_0=0$ and plasma injection between $\rin=\rh$ and  (from left to right) $\rout=2,5,13.5~\rg$. The green, red and blue lines in the bottom panels mark the EM Poynting power, plasma kinetic power and the sum of the two respectively. The screening of $\Epar$ obtained in models $\tt \Theta I$ and $\tt \Theta L$ results in outgoing Poynting flow close to the BZ value. The small drop in the total power at large radii
is due to radiative losses (including IC photons produced below the threshold that are discarded from the simulation).}
\label{fig:energy_flux}
\end{figure}

Figure \ref{fig:energy_flux} exhibits the radial distribution of $\mathbf{D}\cdot \mathbf{B}/B^2$, averaged over the angular direction (top row), 
and (bottom panels) the radial distribution of the energy flow, $\int \it{T^r_{\ t}}dA$, where $\it T^r_{\ t}$
is the total energy flux and $dA$  a surface element of a sphere at radius $r$, in units of the BZ power, here defined as 
\begin{equation}
    \Lbz = \frac{1}{6c} \omega_{\rm H}^2\Phi^2,
\end{equation}
where $\Phi =\int \mathrm{B}^{\it r} \sqrt{\gamma} dA_{\rm H}$  is the magnetic flux on the horizon. The Poynting flow is shown in green, particle energy flow in red and the total power (sum of the two) in blue.
The decrease in Poynting flow seen in model $\tt \Theta S1$ is consistent with the existence of a significant parallel electric field, which exerts work on the pair plasma at the expense of the EM energy. 
The small drop in total power seen in models $\tt \Theta I$ and $\tt \Theta L$ is due to radiative losses. 

\begin{figure}
\includegraphics[width=\linewidth]{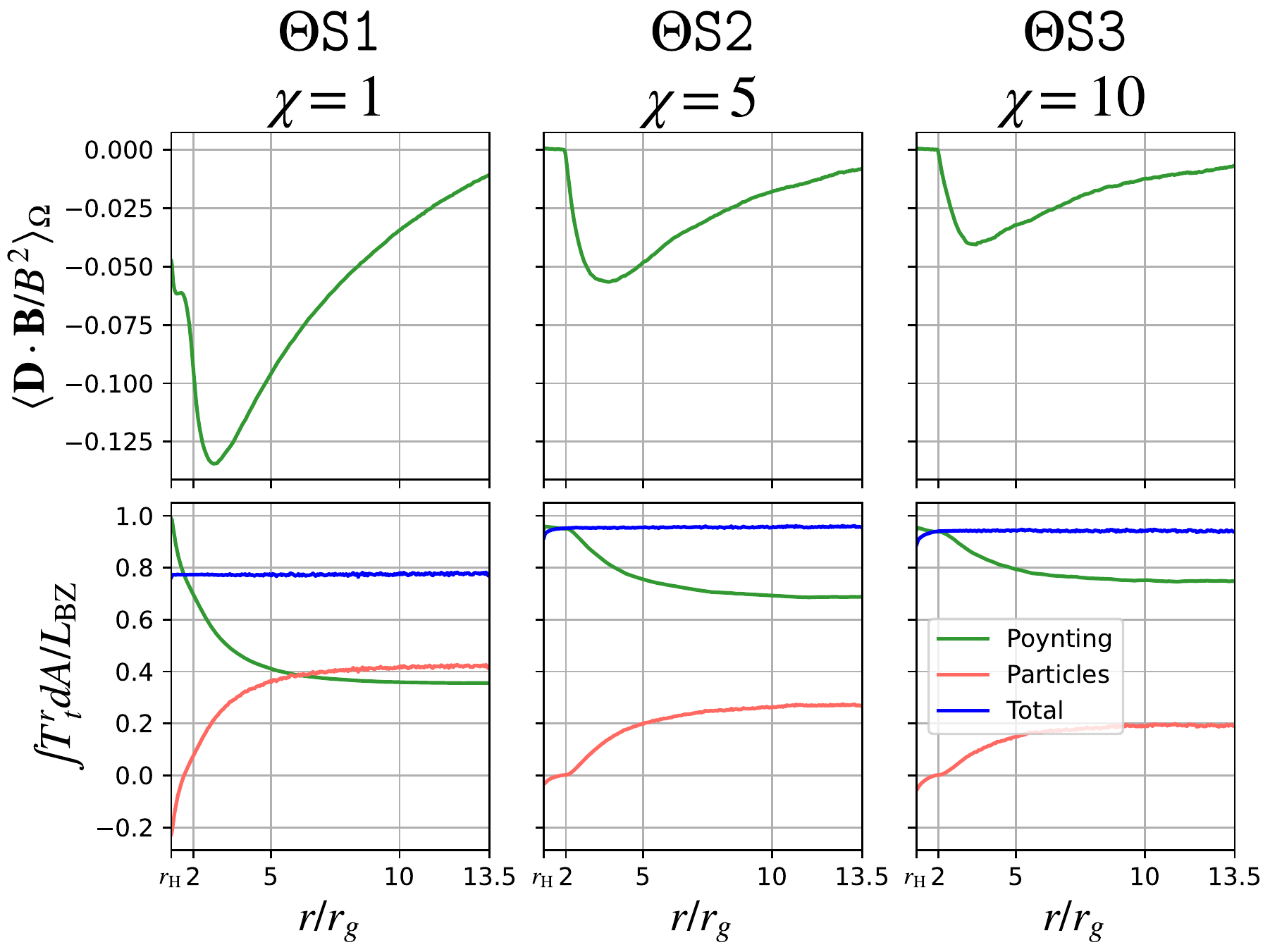}
\caption{
%The radial distribution of the angle-averaged hemisphere parallel electric field, $\Epar=\langle \mathbf{D}\cdot \mathbf{B}/B^2 \rangle_\Omega$ (top), and normalized power, $\int \it{T^r_{\ t}}dA/L_{BZ}$ (bottom) in cases with $\tau=0$, plasma injection between the BH horizon and $\rout=2~\rg$ and (from left to right) $\chi=1,5, 10$. The green, red and blue lines in the bottom panels mark the EM Poynting power, plasma kinetic power and the sum of the two respectively. 
Same as Figure \ref{fig:energy_flux} for a case of injection between $\rin=\rh$ and $\rout=2\rg$, $\tau=0$ and different injection rates with (from left to right) $\chi=1,5, 10$.
The peaks in $\langle \mathbf{D}\cdot \mathbf{B}/B^2 \rangle_\Omega$ (top panels) scale as $\chi^{-1/2}$, implying that to reduce $\Epar$ below $\sim0.01B$ everywhere in the box an injection rate of $\chi >100$ is required.}
\label{fig:energy_flux_model02}

\end{figure}

Increasing the plasma injection rate near the horizon further in model $\tt{\Theta S2}, \tt{\Theta S3}$ improves the screening of $\Epar$,  as seen in Figure \ref{fig:energy_flux_model02}. The figure shows the radial distribution of the solid angle-averaged northern hemisphere parallel electric field, $\langle \mathbf{D}\cdot \mathbf{B}/B^2 \rangle_\Omega$ (top), and total power, $\int \it{T^r_{\ t}}dA$ (bottom), when $\chi$ varies from $\chi=1$ to $\chi=10$. We identify a scaling $\Epar\propto \chi^{-1/2}$ (see figure caption), implying that in order to reduce $\Epar$ below $0.01B$ an injection rate 
of $\chi >100$ is required.

\subsection{External injection}
\begin{figure}
\includegraphics[width=\columnwidth]{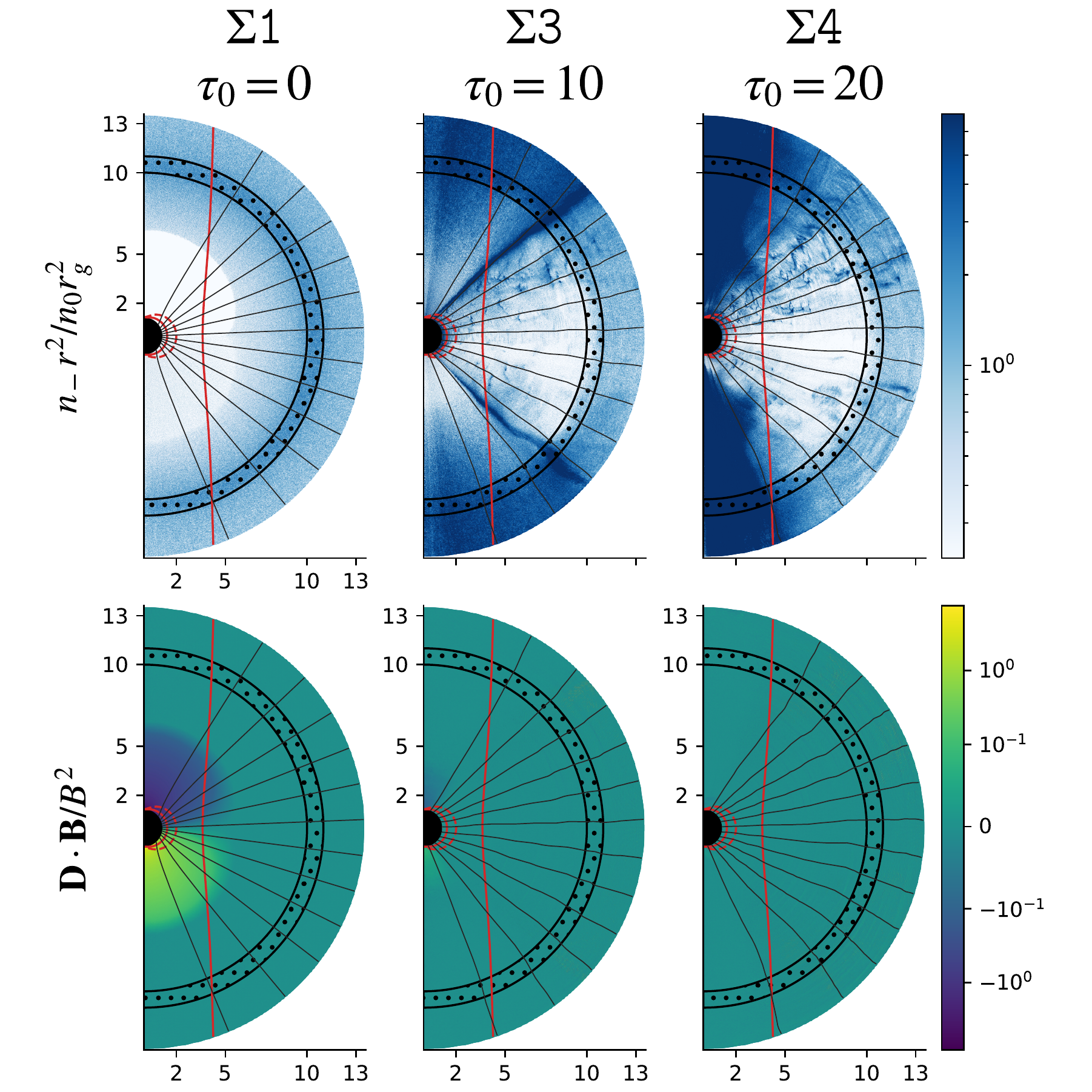}
\caption{Electrons number density (top) and normalized parallel electric field,
$\mathbf{D}\cdot \mathbf{B} /B^2$ (bottom), for cases of plasma injection in a ring with $\rin=10\rg$ and $\rout=11\rg$. Here we set the initial optical depth for pair creation to be (from left to right) $\tau_0=0, 10, 20$. The injection zones are marked with black dots, magnetic field lines with gray solid lines and the outer light cylinder with a dashed red line. screening is obtained in the two right panels with $\tau_0\geq10$.}
\label{fig:electric_field2}
\end{figure}

\begin{figure}
\includegraphics[width=\linewidth]{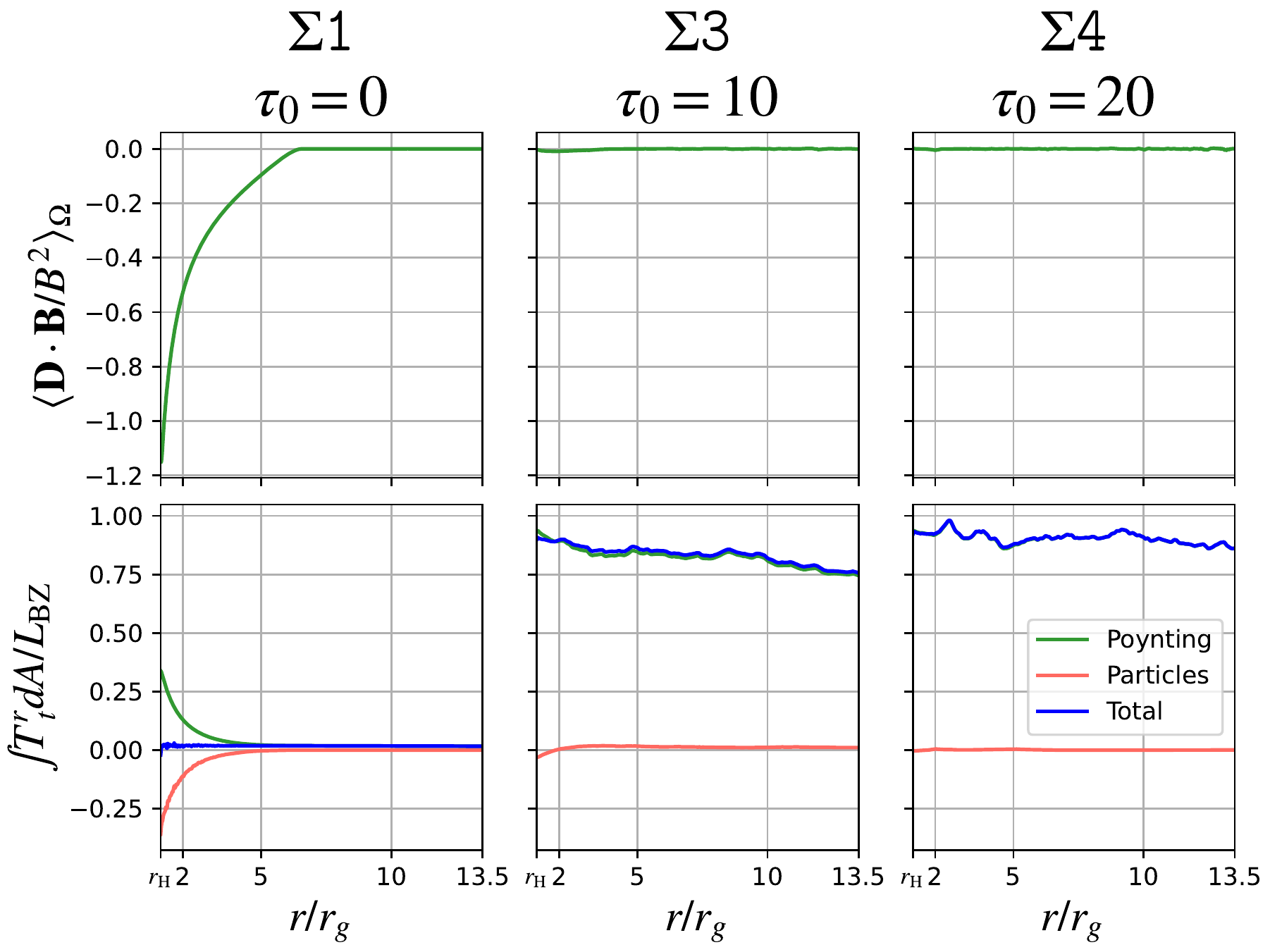}
\caption{The radial distribution of the solid angle-averaged northern hemisphere parallel electric field, $\langle \mathbf{D}\cdot \mathbf{B}/B^2 \rangle_\Omega$ (top), and power, $\int \it{T^r_{\ t}}dA/\Lbz$ (bottom) for cases of plasma injection in a ring with $\rin=10\rg$, $\rout=11\rg$ and varied optical depth, where from left to right, $\tau_0=0, 10, 20$. The green, red and blue lines in the bottom panels mark the EM Poynting power, plasma kinetic power and the sum of the two respectively. In the models with high opacity better screening is obtained resulting in an 
outgoing Poynting flow close to the BZ value. The drop in the total power at large radii is due to radiative losses (including IC photons produced below the threshold that are discarded from the simulation).}
\label{fig:energy_flux2}
\end{figure}

In the second suite of experiments we injected  plasma in a ring between $\rin =10 \rg$ and $\rout=11 \rg$ (shown schematically in the middle panel 
in Fig. \ref{fig:models}), varying the pair injection rate 
$\chi$ and the fiducial optical depth for pair creation, $\tau_0$, between the different runs. 
Snapshots from simulations with $\chi=1$ and $\tau_0=0, 5, 10, 20$, taken at times after the system (in each run) has reached a steady state, are
exhibited in Figure \ref{fig:electric_field2}.   The top panel delineates the normalized electron density and the bottom panel shows $\mathbf{D}\cdot \mathbf{B}/B^2$, as
in Fig \ref{fig:density_Epar_inner}.  As seen, when pair creation is switched off ($\tau_0=0$, model $\tt \Sigma1$) the injected plasma 
is unable to screen the entire magnetosphere, even though the plasma density in the injection ring and its vicinity exceeds the GJ density considerably. 
A large vacuum gap pertains in the inner region, within about $5\rg$.  Inside the gap electrons are accelerated by the field aligned electric field $\Epar$ 
inwards in the southern hemisphere and likewise positrons in the northern hemisphere.  The supply of plasma into the ergosphere by the accelerated pairs 
induces electric current that generates an outward Poynting flow (Fig \ref{fig:energy_flux2}).  However, the outflowing Poynting energy is compensated 
by the inflowing energy carried by the inwards moving pairs.  The net positive energy flux is small, about $0.02 \Lbz$. 

Switching on the interaction with the ambient soft photons gives rise to prodigious generation of gamma rays and newly created pairs for large enough $\tau_0$, as expected.
We find that complete screening of the entire magnetosphere occurs at $\tau_0 \gtrsim 20$ (model $\tt \Sigma4$).  The total energy flux is carried completely by the Poynting flow, and approaches its maximum value.  We also observe that a small fraction (a few percents) of the energy flux emerging from ergosphere is converted to intermittent 
(high-energy) radiation (curvature radiation through radiation back-reaction and IC photons below the pair creation threshold).  
Note that unlike IC photons above the pair production threshold, curvature photons and IC photons below the threshold are not treated as 
PIC particles in the simulations, and are not included in the plot of the radiation energy flux in the figures.
The overall behaviour of the system is similar to that 
presented in \cite{Crinquand2020}, except for the density distribution
which in our case is partly imposed by the external plasma injection process.  

One might suspect that the formation of a macroscopic vacuum gap in the case of $\tau_0=0$ is a consequence of insufficient plasma supply, and that 
increasing the injection rate sufficiently might  ultimately result in a complete screening.  To examine how the magnetosphere responds to increased 
plasma injection rate, we performed simulations with $\tau_0=0$, $\chi=30$ (model $\tt \Sigma 6$) and $\chi= 50, 100$ (these models are not listed in table \ref{tab:models}).  
Interestingly, we find a cyclic dynamics for $\chi> 10$\footnote{For $\chi=10$ (model $\tt \Sigma 5$) we observe damped oscillations that tend to converge to 
a state with a starved inner region (between the horizon and $r\approx 3\rg$). This seems to be a transition case between steady and oscillatory solutions.}.   The inner gap exhibits oscillations with a period of about $70 \tg$, during which the gap 
size repeatedly shrinks to a minimum (at which it extends from the horizon to some radius within the ergosphere) and then expands to a maximum size in 
excess of $5 \rg$ (a link to the movie showing this behaviour is given in table \ref{tab:models}, model $\tt \Sigma6$).
The density in the region outside the injection ring exhibits strong time modulations that correlate with the gap activity.  
For $\chi=100$ the density at maximum largely exceeds $n_{\rm GJ}$ in most of the simulation box, approaching a few houndreds $n_{\rm GJ}$ in the injection zone.  
Within our limited dynamic range, we find no evidence for a tendency of the system to reach a steady state as $\chi$ is increased.

To examine the dependence on the width of the injection ring we ran a simulation with hot plasma injection into a ring extending from $\rin=9 \rg$
to the outer edge of the simulation box, $\rout=13 \rg$ (not listed in table \ref{tab:models}).  We find cyclic dynamics, very similar to that described above. 
A similar behaviour is also exhibited in the cases with a torus configuration (see Sec. \ref{sec:torus} below).  We conclude that this quasi-cyclic evolution 
occurs in cases where plasma is injected outside the outer light cylinder.

The following heuristic argument offers an explanation for this behaviour: When the magnetosphere is nearly completely screened, and a BZ outflow is
established, a stagnation surface forms across which which the velocity of injected plasma changes sign \citep{globus2014}.  This double flow structure 
is a consequence of the causal structure of the magnetosphere. 
In particular, plasma within the inner light surface must be flowing inwards and plasma above the outer light surface must be flowing outwards. 
This implies that plasma must be continuously injected between the inner and outer light surfaces to keep the outflow in a force-free state at all times.
Now, in the simulations described above plasma is injected only above the outer light surface, and since this plasma cannot reach the region below the stagnation
surface, over time it becomes devoid of plasma and a macroscopic gap forms.  If the injection rate is not high enough, as in the cases with $\chi \lesssim 10$, 
a steady state is established, in which part of the injected plasma is flowing outwards, and part is being pulled into the BH by the parallel electric 
 field generated in the starved magnetospheric region around the BH.  When the injection rate is high enough, as in the runs with $\chi >10$, enough plasma 
 is being pulled inwards during phases of magnetospheric starvation to nearly screen the entire magnetosphere.  A BZ outflow is then formed for a time it takes 
 the plasma below the stagnation surface to be evacuated, leading again to formation of a large vacuum gap in the inner region and the cycle repeats. 

\subsection{Torus configurations}\label{sec:torus}
In our final suite of experiments we inject hot plasma ($\kb T/\me c^2 =10$) into a ring 
sector with an opening angle $\Delta \theta$ about the equatorial plane (right panel in Fig. \ref{fig:models}), 
located between radii $\rin=10\rg$ and  $\rout=11\rg$  (that is, the ring
extends from $\tmin=\frac{\pi-\Delta\theta}{2}$ to $\tmax=\frac{\pi+\Delta\theta}{2}$). 
In these runs the entire injection region is located outside the outer light cylinder, and numerical effects that might be associated with
injection near the axis are avoided.
As in the other cases, the density outside the injection zone is taken to be zero initially.  
We examined cases with $\Delta\theta = 60^\circ$ (models $\tt{\Xi N1} - \tt{\Xi N3}$) and $\Delta\theta = 120^\circ$ (models $\tt{\Xi W1} - \tt{\Xi W3}$).  
We find a similar behaviour to the previous cases; at $\chi = $ a few, the system 
reaches a quasi steady-state at $t \approx 70 \tg$.  At higher injection rates (particularly for the $\Delta\theta=120^\circ$ case) the system exhibits 
cyclic oscillations similar to those seen in the full rings with $\chi>10$.   In all cases the plasma is confined to the magnetic field lines, 
as expected for $\sigma\gg1$; the polar regions 
at $\theta <\tmin$ and $ \theta>\tmax$ remains evacuated from charges for the entire simulation.   
The net energy flux emerging from the ergosphere is small (practically zero for $\Delta\theta =60^\circ$ and $\sim 0.1$ for $\Delta\theta =120^\circ$).
When the interaction with external radiation is 
switched on ($\tau_0\ne0$), photons produced through IC scattering inside the ring section slowly leak out, producing new pairs, whereupon 
the entire magnetosphere is eventually filled with plasma and screened, and the extracted power approaches $\Lbz$. 

\begin{figure}
\includegraphics[width= \linewidth]{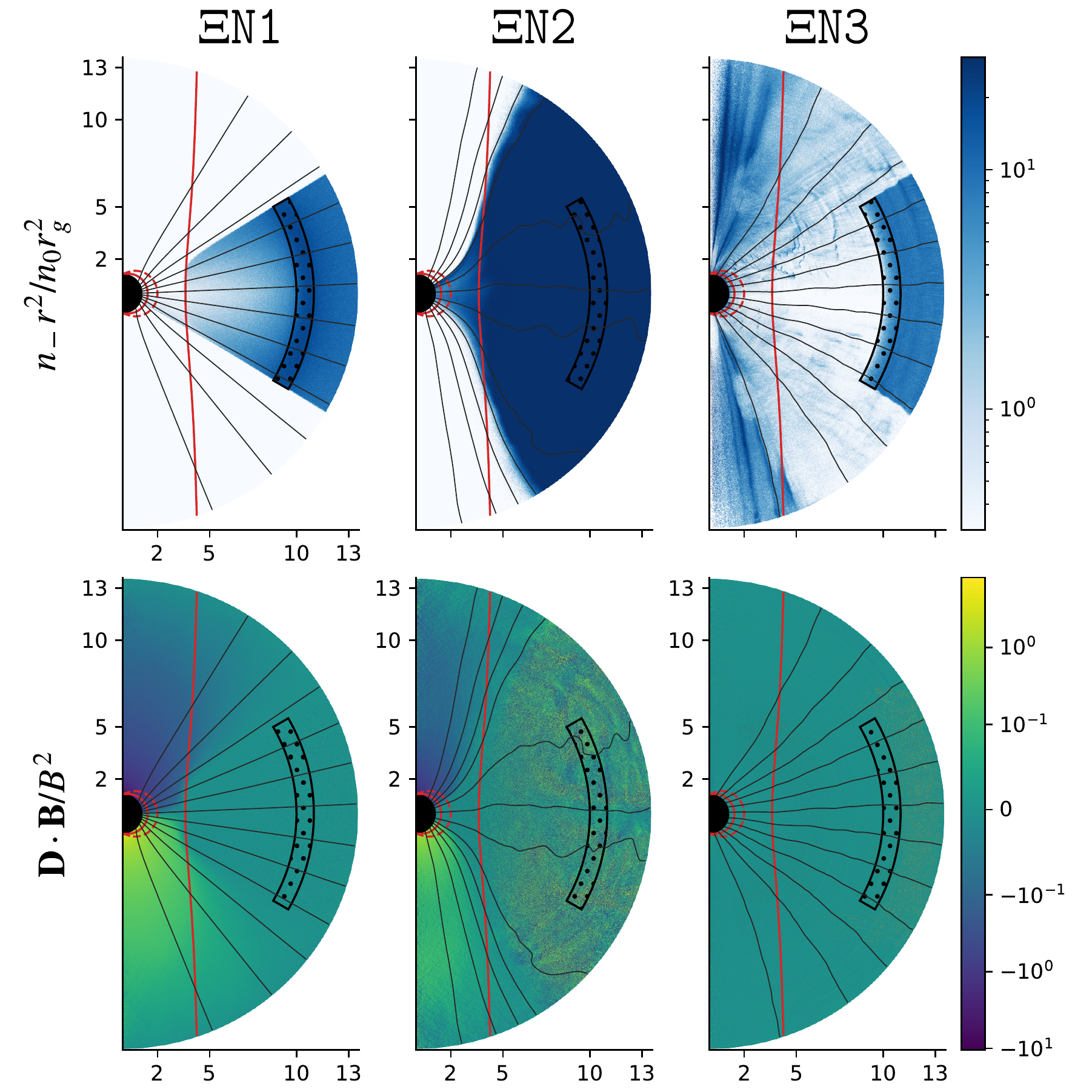}
\caption{Same as Fig. \ref{fig:electric_field2} for injection in a narrow ring section, 
with high fiducial magnetization and $\tau_0=0$  (model $\tt\Xi N1$), $\tau_0=20$ (model $\tt\Xi N3$), and low  fiducial magnetization 
with $\tau_0=0$ (model $\tt \Xi N2$).  In runs $\tt{\Xi W1}, \tt{\Xi N3}$ the magnetization largely exceeds unity everywhere inside the 
simulation box at all times.  In run $\tt\Xi N2$ the injection zone is overloaded with plasma ($\sigma(r_{\rm{inj}}) <1$) and it is 
seen that the hot plasma can diffuse across field lines. However, the BH outflow is completely chocked (see text for details).
\label{fig:sigma1} }
\end{figure}

We also ran two cases for each configuration with fiducial magnetization $\sigma_0 =5\times 10^3$, one with low injection rate, $\chi=1$, 
and one with high injection rate (models $\tt \Xi N2$ and $\tt \Xi W2$).
The actual magnetization in the injection zone is around unity for the low injection rate cases and below unity in the high injection cases. 
We find a strong distortion of magnetic field lines and production of waves, as naively expected.  Plasma from the injection zone
diffuses into part of the polar region; in the case with high injection rate (see model $\tt \Xi N2$ in Fig. \ref{fig:sigma1} for example)
it penetrates down to an angle of about $15^\circ$ in 
the northern hemisphere ($165^\circ$ in the southern one).  Close to the poles ($\theta< 15^\circ$) the density remains very low (nearly zero).  
We find an emerging Poynting flux from the horizon, mainly within the injection section, but it decays over a few $\rg$, transferring energy 
to particles. It seems that this energy is given back to the torus.  This choking of BH outflow is anticipated on overloaded 
field lines \citep{globus2014}. In the polar region, where the plasma density is low and the 
magnetization is high ($\gg1$), the power of the emerging Poynting flow is very small.

\begin{figure}
\includegraphics[width= \linewidth]{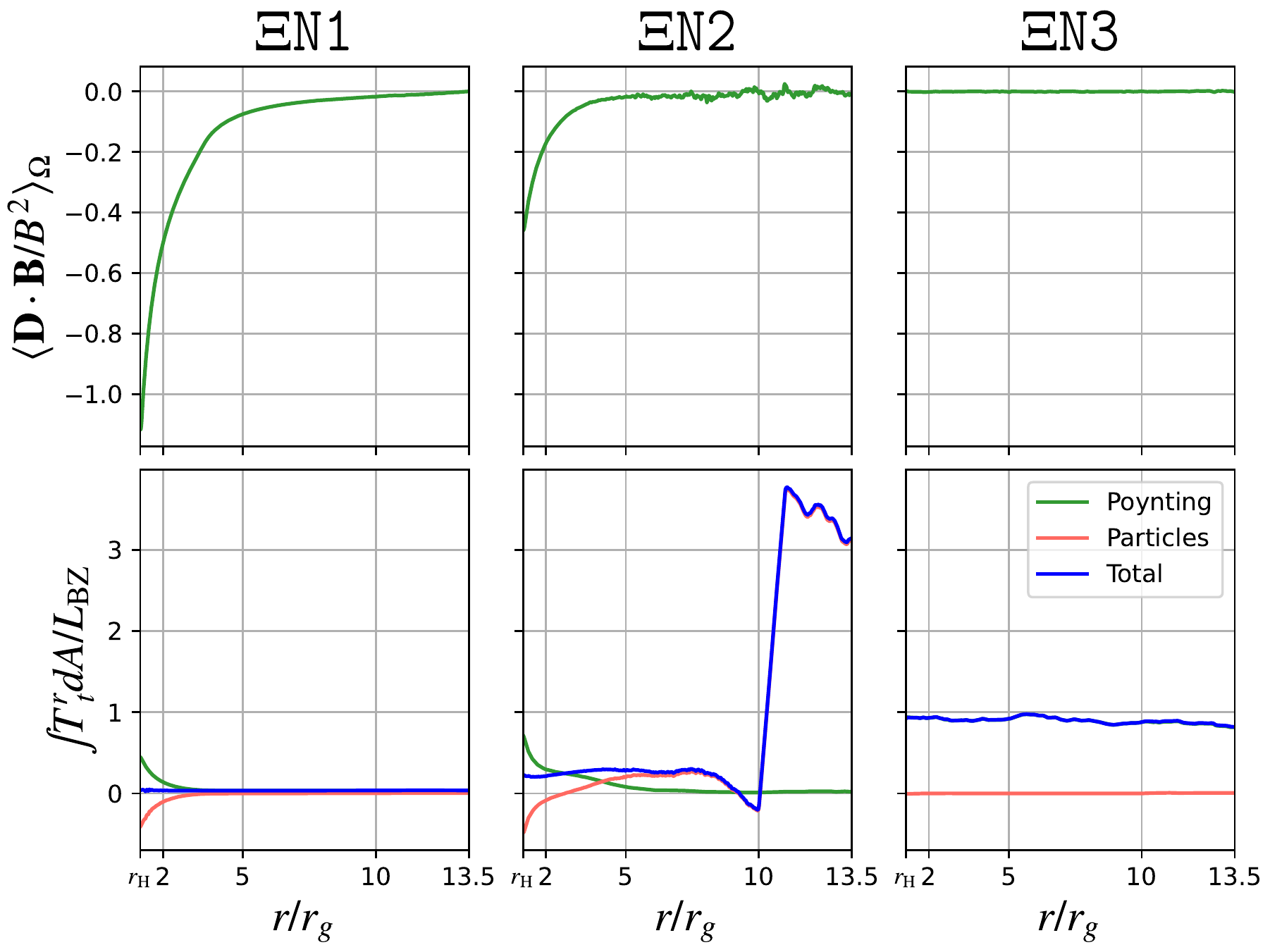}
\caption{Same as Fig. \ref{fig:energy_flux2} for the runs shown in Fig. \ref{fig:sigma1} 
\label{fig:sigma1_flux} }
\end{figure}
\section{Conclusion}
We studied the response of a BH magnetosphere to plasma injection by means of radiative 2D GRPIC simulations, that incorporate photon 
generation and pair production through interactions with a given radiation field (representing disk emission) in a self-consistent manner.  
We conducted several sets of numerical experiments in which relativistically hot plasma is injected locally at a prescribed rate in a given section of the magnetosphere,
varying the geometry of the injection zone, the injection rate and the intensity of ambient radiation field between the different experiments. 
In all of the experiments a monopole magnetic field configuration was adopted in the initial state.

We find that when the interaction of pairs with the external radiation field is switched off (formally, setting the intensity to zero),
injection of hot plasma can completely screen the magnetosphere, provided the injection zone is located within the outer light cylinder and 
the injection rate is high enough.  In that case we observe the formation of a Poynting flow that emanates from the BH horizon and propagates to infinity 
with nearly maximum BZ power.  On the other hand, when the plasma is injected beyond the outer light cylinder complete screening never occurs; at modest 
injection rates the system reaches a steady-state, with a macroscopic vacuum gap extending from the vicinity of the horizon up to the outer light cylinder roughly.
At higher injection rates the magnetosphere exhibits cyclic dynamics, during which it oscillates between nearly complete screening to extended starvation. 

In all cases, when the interaction with the external radiation field is switched on, and the opacity is large enough ($\tau_0\gtrsim20$),
complete screening always ensues, with nearly maximal energy extraction.  In the cases where the plasma is injected externally beyond the outer light cylinder,
we find that a fraction of a few percents of the extracted energy (the maximum BZ power) is converted to VHE radiation through IC emission and radiation backreaction (curvature
emission), as found earlier in \citet{Crinquand2020}. 
%\benjamin{I find the $\Theta W 3$ case interesting: even though plasma is injected far from the BH, it doesn't matter and the magnetosphere is still screened. So the necessary condition is that photons can trigger a pair cascade close to the BH, and once it the cascade is on, the plasma supply becomes irrelevant?}

Our main conclusion is that, in reality, sporadic injection of plasma from the accretion flow into the polar region by some (yet unspecified) 
process, is unlikely to screen the magnetosphere completely at all times, and prevent intermittent sparking.  Formation of spark gaps during 
charge starvation episodes should lead to variable TeV emission with a luminosity that can approach a few percents of the jet power, as
proposed earlier \citep[e.g.,][]{Levinson2000,Neronov2007,Levinson2011,Hirotani2016}.

%\benoit{More comments: (i) Would be useful to say where the Bondi radius is for the temperature of the injected plasma, (ii) Fig 7, it seems that the shaded zone does not correspond to the geometry of the injection zone, correct? (iii) Should we show the equivalent of Fig 6 for the torus case for completeness?}

\section{Data Availability}
The data underlying this article will be shared on reasonable request to the corresponding author.

\section{Acknowledgments}
AL acknowledges support by the Israel Science Foundation grant 1995/21. This project has received funding from the European Research Council (ERC) under the European Union’s Horizon 2020 research and innovation program (Grant Agreement No. 863412). This research was facilitated by the Multimessenger Plasma Physics Center (MPPC), NSF grant PHY-2206607.

\bibliographystyle{mnras}

\bibliography{paper}

\end{document}